\documentclass[11pt]{article}

\pdfoutput=1
\usepackage{setspace}
\usepackage{amssymb}
\usepackage{bm}
\usepackage{amsmath}
\usepackage{amsmath, amsthm, amssymb}
\usepackage[pdftex]{graphicx}
\usepackage{natbib}
\usepackage{lineno}
\usepackage{lscape}
\usepackage{amsfonts}
\usepackage[figuresright]{rotating}

\usepackage{todonotes}

\newcommand{\sgn}{\text{sgn}}

\newcommand{\beq}{\begin{equation}}
\newcommand{\eeq}{\end{equation}}
\newcommand{\bed}{\begin{displaymath}}
\newcommand{\eed}{\end{displaymath}}

\newcommand{\tr}{^{T}}
\newcommand{\bbeta}{\boldsymbol{\beta}}

\newcommand*{\subscript}[1]{\ensuremath{_\textrm{{\scriptsize #1}}}}

\def \bx {{\mathbf x}}

\def \bzero {{\mathbf 0}}

\begin{document}
\title{Scale estimation and data-driven tuning constant selection for M-quantile regression}
\author{James Dawber \\
Southampton Statistical Sciences Research Institute\\  University of Southampton, UK
\and
Nicola Salvati\\
Dipartimento di Economia e Management \\ Universit\`a di Pisa, Italy
\and Timo Schmid \\
 Institute of Statistics and Econometrics\\ Freie Universit\"at Berlin, Germany
\and Nikos Tzavidis\\
Southampton Statistical Sciences Research Institute\\  University of Southampton, UK
} \date{} \maketitle

\begin{abstract}
M-quantile regression is a general form of quantile-like regression which usually utilises the Huber influence function and corresponding tuning constant. Estimation requires a nuisance scale parameter to ensure the M-quantile estimates are scale invariant, with several scale estimators having previously been proposed. In this paper we assess these scale estimators and evaluate their suitability, as well as proposing a new scale estimator based on the method of moments. Further, we present two approaches for estimating data-driven tuning constant selection for M-quantile regression. The tuning constants are obtained by i) minimising the estimated asymptotic variance of the regression parameters and ii) utilising an inverse M-quantile function to reduce the effect of outlying observations. We investigate whether data-driven tuning constants, as opposed to the usual fixed constant, for instance, at c=1.345, can improve the efficiency of the estimators of M-quantile regression parameters. The performance of the data-driven tuning constant is investigated in different scenarios using model-based simulations. Finally, we illustrate the proposed methods using a European Union Statistics on Income and Living Conditions data set.
\end{abstract}


\section{Introduction}\label{intro}
Regression models are commonly used to estimate the mean of a random variable conditional on a set of covariates. Inference is then made based on distributional assumptions and independence. In practice, these assumptions are sometimes not met, for example in the presence of outliers. Robust regression models were developed to make regression possible when the conditional distribution is atypical, such as with heavy-tailed distributions. \citet{Hub64} developed a versatile approach to robust estimation called M-estimation which uses a loss function $\rho$ or its derivative the influence function $\psi$. Regression M-estimates can also be found by either minimising the loss function or if this function is differentiable then equating the influence function to zero and solving \citep{Hub73}. Let $y$ be a response variable of interest with $p$ covariates $\mathbf{x}$ that can include an intercept, then a general approach to estimate the regression parameter vector $\bbeta$ is to solve:
\begin{equation} \label{eq:mq}
\sum_{i=1}^n\psi\left(y_i-\bx_i\tr\hat{\bbeta}\right)\bx_i=\bzero
\end{equation}
where $\psi$ is an influence function usually assumed to be bounded and monotone non-decreasing over the real line with $\psi(0)=0$. The well-known least squares regression is the case when $\psi$ is the identity function. Robust regression makes use of influence functions that are bounded, reducing the influence of outlying residuals. A commonly used robust influence function is the Huber influence function:
\begin{equation}\label{eq:psi_huber}
\psi_{c}(u) = \left\{\begin{array}{lc} u & |u|\leqslant c \\
               c~ \sgn(u) & |u| > c, \\
\end{array}\right.
\end{equation}
where $c>0$ is a predetermined tuning constant. Any observations that exceed the intervals set by $c$ are down-weighted, which makes the estimates more resistant to outliers. In this context, these down-weighted observations are said to be `Huberised'. Notice that when $c\rightarrow \infty$ the Huber influence function is equivalent to estimating the mean, and when $c\rightarrow 0$ it is equivalent to estimating the median. Hence this Huber estimator acts as a generalisation of these two common special cases.

Two noteworthy challenges arise when performing regression using the Huber influence function, the first is that a value for $c$ must now be selected, and secondly the estimates will no longer be scale invariant, hence a nuisance scale parameter ($\sigma$) must be introduced and estimated. Hence, Huber regression estimates of $\bbeta$ are found through solving the following estimating equations for $\hat{\bbeta}$:
\begin{equation} \label{eq:mq1}
\sum_{i=1}^n\psi_c\left(\frac{y_i-\bx_i\tr\hat\bbeta}{\hat\sigma}\right)\bx_i=\bzero, 
\end{equation}
where $\hat{\sigma}$ is an estimator of $\sigma$. The trade-off for using this robust approach to regression is that appropriate $c$ and $\hat\sigma$ must now be included in the model. In practice, it is common for a default value of $c=1.345$ to be used, as first suggested by \citet{Hol77}, which ensures $95\%$ efficiency for normal residuals, and if more or less robustness is required then a different tuning constant can be selected. \citet{Wan07} proposed a data-driven approach to automatically select the value of $c$ such that the asymptotic efficiency of the estimate is maximised. \citet{Hub64} also suggested a similar approach involving efficiency.

To estimate $\sigma$, \citet{Hub64} first proposed a scale estimator such that the second moment of the estimator should be consistent under normality, however more commonly in practice a simpler median absolute deviation (MAD) scale estimator is used to estimate $\sigma$. The MAD is a highly robust scale estimator and as such is often the default choice. There are many other potential robust estimators of scale, as discussed by \citet{Rou93}.

Robust regression, just like the traditional regression model, provides a rather incomplete picture of the behaviour of the response variable given the covariates, in much the same way that the median and the mean give an incomplete picture of a distribution. In such cases, it may be preferable to fit an ensemble of regression models, each one summarising the behaviour of a different percentage point, or quantile, $q \in (0,1)$, of the response variable conditional on the covariates. Thus giving a more complete picture of the entire conditional distribution. \citet{Bre88} proposed M-quantile (hereafter, MQ) regression models defined by a `quantile-like' generalisation of quantile regression \citep{Koe78} and the lesser known expectile regression \citep{New87} based on influence functions.

Let the $q$-th MQ be denoted as $Q_q$ and the MQ linear regression model defined as $Q_q(\mathbf{x}_{i})=\mathbf{x}^T_{i}\bbeta_q$ with $q \in (0,1)$. Then estimates of the MQ regression parameters can be made by solving the following estimating equations for $\hat\bbeta_q$:
\begin{equation} \label{eq:esteq}
\sum_{i=1}^n\psi_{q}\left(\frac{y_i-\bx_i\tr\hat\bbeta_q}{\hat{\sigma}_{q}}\right)\bx_i=\bzero,
\end{equation}
where $\hat{\sigma}_{q}$ is an estimator of a scale parameter $\sigma_q$ and the MQ influence function is:
\begin{equation} \label{eq:mq2}
\psi_{q}(u) = \left\{\begin{array}{lc} 2(1-q)\psi(u) & u\leqslant 0 \\
               2q\psi(u) & u > 0, \\
\end{array}\right.
\end{equation}
and the other influence function $\psi$ nested within, is selected as before in Equation~(\ref{eq:mq}). The parameter $\sigma_q$ is a suitable measure of the scale which ensures the MQ estimator is scale invariant. If $\psi(u)=u$ the expectile of order $q$ is obtained, which represents a quantile-like generalisation of the mean, while the quantiles are obtained if $\psi(u)=\sgn(u)$. Also, note that setting $q=0.5$ will provide an estimator of the mean or median respectively, highlighting how this MQ estimator is a generalised form of these wider-known estimators. In the specific cases of the expectile and the quantile the parameter $\sigma_q$ is unnecessary as these estimators are scale invariant. Although quantiles and expectiles are most commonly used in practice, MQ estimators with other nested influence functions can be used, such as the Huber influence function $\psi_c$, which provides a tuning constant with the quantile $c\rightarrow 0$ and expectile $c\rightarrow \infty$ at two extremes. We refer to the MQ estimator using the Huber influence function as the Huber MQ.

There are a vast range of applications in which quantile, expectile and Huber MQ regression have all been applied. We refer to \citet{Wal15} for an introduction to quantile and expectile regression as well as applications, which for expectiles, are often used for measuring risk in actuarial science \citep{Bel14, Dao18}. Huber MQs are not as widely used as the more interpretable quantiles, but they combine the robustness of quantiles with the efficiency of expectiles into a single framework determined by a single tuning constant. Huber MQs can be preferable to expectiles in the presence of outliers, where the robustness properties improve estimation. The primary disadvantage of quantile regression is that the maximum likelihood solution using influence functions is not unique, a problem heightened by small sample sizes. Unlike expectiles and Huber MQs that can be solved using iteratively re-weighted least squares (IRLS), quantile regression estimates are typically found through minimising the loss function using linear programming \citep{Koe78}. Furthermore, quantile estimation is inefficient when residual distributions are close to normal. Due to these limitations for both expectile and quantile regression, Huber MQs have been used in certain applications that require robustness, guaranteed uniqueness, and higher efficiency such as when data are few, or when trading robustness for efficiency via a tuning constant is desirable. Hereafter, for simplicity, we refer to the Huber MQ estimator merely as the MQ estimator.

The most prominent field of application for MQ regression is in small area estimation (SAE), see \citet{Daw19} for an overview of these methods. \citet{Cha06} showed that MQ regression models perform comparably to traditional SAE models using mixed models, and perform better when outliers are present in the data. Similarly to SAE methods, MQ regression models have been used in poverty mapping \citep{Tza08, Giu09}. MQs were also used in estimating acidity in north-eastern US lakes \citep{Pra08}, and in an analysis of temporal gene expression data \citep{Vin09}. Also, MQ random-effects models were introduced by \citet{Tza15} and \citet{Borgonietal2016} with applications to longitudinal data. \citet{Del19} adapted this MQ random-effects model and applied it to air quality data, also \citet{Cha19} introduced MQ modelling to data linkage. As well as applications, theoretical developments of MQ models have also been made, most notably recently by \citet{Bia15}, \citet{Bia18} and \citet{Alf17}.

The disadvantages of MQs compared to quantiles and expectiles mirror that of the Huber estimator compared to the median and mean, that is, the requirement for $c$ to be selected, as well as a nuisance scale parameter $\sigma_q$ that needs to be estimated to ensure scale invariance. Solutions to these two problems have been solved in very similar ways. The tuning constant is commonly suggested to be $c=1.345$ for all $q$, despite having $95\%$ efficiency only when $q=0.5$, and the scale estimator is commonly suggested to be based on the MAD as first proposed by \citet{Cha06}:
\begin{equation}\label{eq:nMAD}
 \hat{\sigma}_q=med_i(|\hat\epsilon_{iq}|)/\Phi^{-1}(3/4),
 \end{equation}
with $med_i(\cdot)$ returning the median for all observations $i$, and $\hat\epsilon_{iq} = y_i-\bx_i\tr\hat\bbeta_q$. The constant $\Phi^{-1}(3/4)$, where $\Phi(\cdot)$ is the distribution function of the standard normal distribution, ensures consistency with the standard deviation when residuals are normal and $q=0.5$. The majority of MQ applications refer to this MAD estimator as the suggested approach. An iterative method is needed to obtain an estimate of $\sigma_q$, as it is for solving the full regression in Equation~(\ref{eq:esteq}). An IRLS algorithm or the Newton-Raphson algorithm can be used \citep{Bia18}.

With a growing number of MQ regression applications in the literature it seemed appropriate to revisit and inspect the suitability of the pre-existing suggestions for selecting $c$ and estimating $\sigma_q$, as these are key weaknesses to MQ models. The choice of both these components is very important for effectively using MQ models, and they also have an interrelated relationship. This is because the scale parameter also has a role in determining which observations are Huberised by the tuning constant $c$. Since the role of the scale parameter is to standardise the observations, it affects which observations fall outside the tuning constant threshold and are consequently Huberised. More explicitly, residuals are Huberised when:
\begin{equation}\label{eq:sigandk0}
\left|\hat\epsilon_{iq}\right| > c\hat{\sigma}_{q},
\end{equation}
hence it is important that $\sigma_q$ is sensibly chosen and estimated to allow for appropriate Huberising by $c$. Due to this interrelationship between $c$ and $\sigma_q$, it seems sensible to select the scale estimator and tuning constant in conjunction, to help achieve the optimal level of robustness.

The primary aims of this paper are twofold. First, to assess and explore different scale estimators for MQs and determine which of these estimators is best for general use. Second, to explore the role of tuning constant selection beyond $q=0.5$ and introduce two data-driven approaches to selecting MQ tuning constants. Through assessing the role of tuning constant and scale estimators on MQ estimation, improved understanding and confidence can be gained in the use of MQ regression models, as well as highlighting distribution types and problems where certain tuning constants and scale estimators are inappropriate. 

Four different MQ scale estimators will be evaluated including the variant of the MAD shown in Equation~\eqref{eq:nMAD}, also the more complete MAD estimator outlined in \citet{Bia15}. Thirdly, a maximum likelihood estimator approach introduced by \citet{Bia18} based on a parametric distribution associated with the Huber MQ loss function called an asymmetric least informative (ALI) distribution. Finally, a method of moments approach based on this ALI distribution is introduced. These scale estimators are described in more detail in Section~\ref{sec:2}, and their performance and appropriateness compared in a simulation study in Section~\ref{sec:scalecomp}.

Following this we present two methods for selecting MQ tuning constants, one for a specific or local $q$, the other for a global $q$. The first method, presented in Section \ref{sec:3}, follows the approach by \citet{Wan07} except different values of the tuning constant are obtained for each $q$. The second method selects a global value of $c$ based on an assumed contaminated normal distribution response which identifies the tuning constant that best down-weights the contaminated observations. A global approach is important since in many applications one single value of $c$ is selected over an ensemble of differing values of $q$. In Section \ref{sec:sim} the two approaches are evaluated through a simulation experiment. Section \ref{sec:cbs} presents the application of the two methods for the choice of the tuning constant to a European Union Statistics on Income and Living Conditions (EU-SILC) data set. Finally, in Section \ref{sec:final}, we summarise our main findings and emphasise the implications for future MQ regression applications.

\section{M-quantile scale estimators}\label{sec:2}
In this section we explore four different approaches to MQ scale estimation for continuous data, three previously introduced to the literature based on MAD and maximum likelihood, and one new proposal based on the method of moments. These four scale approaches are first individually described and their attributes detailed, before their differences in practice are inspected via a simulation study and application to real data.

\subsection{Median absolute deviation approach}
The MAD approach is a widely used method of estimating $\sigma_{q}$ in MQ estimation. This is because it is suggested in \citet{Cha06} which is one of the earliest applications of MQ regression, where the scale estimator is calculated using Equation~\eqref{eq:nMAD}. However this is a simplified variant of the MAD, the correct MAD estimator should be defined as:
\begin{equation}\label{eq:cMAD}
 \hat{\sigma}=med_i (|\hat\epsilon_{iq} - med_{j}(\hat\epsilon_{jq})|)/\Phi^{-1}(3/4).
 \end{equation} 
We distinguish the two MAD estimators by referring to the estimator defined by Equation~\eqref{eq:nMAD} as the `naive' MAD (nMAD) estimator, and the  estimator defined by Equation~\eqref{eq:cMAD} as the `corrected' MAD (cMAD) estimator. The term `naive' is used because of the assumption that $med_{j}(\epsilon_{jq})=0$, which is a natural assumption when $q=0.5$ and with symmetric distributions. But residuals for an MQ estimate with extreme $q$ should not be symmetric about the location estimate, hence this assumption is not generally sensible. Not making this assumption leads to a `corrected' MAD estimator with this additional term which reflects the traditional definition of the MAD estimator. Nevertheless, the nMAD estimator still qualifies as a viable scale estimator as it is invariant to scale, sign and location shifts.

Scale estimators can be expressed similarly to location $M$-estimators because they can be expressed as solutions to estimating equations which use influence functions. So in a regression case with $p$ parameters in the $\bbeta_{q}$ vector, the additional scale parameter makes a total of $p+1$ parameters. In this case the influence function is a $p+1$ vector which for the nMAD estimator can be expressed as:
\begin{equation*}
    \bm{\psi_{\subscript{nMAD}}}(y_i, \bm{x}_i; \bm{\beta}_{q}, \sigma_{q})=
\begin{pmatrix}
\psi_q\left(\frac{y_i - \bm{x}_i^T\bm{\beta_q}}{\sigma_q}\right)\bm{x_{i}}  \\ \text{sign}\left\{\left| \frac{y_i - \bm{x}_i^T\bm{\beta}_q}{\sigma_q} \right| - \Phi^{-1}(3/4) \right\}
\end{pmatrix}.
\end{equation*}
The estimates are then found by solving $n^{-1}\sum_{i=1}^n\bm{\psi_{\subscript{nMAD}}}(y_i, \bm{x}_i; \hat{\bm{\beta}}_q, \hat{\sigma}_q) = \bm{0}$. Estimating the scale parameter using these estimating equations is equivalent to using Equation~\eqref{eq:nMAD}. The benefit of expressing the scale estimator using influence functions is that it provides a simple way to ensure the estimator is robust to outliers by checking the function is bounded. In this case the sign function clearly ensures this.

In the same way, the influence function vector for the MQ estimator using the cMAD scale estimator can be shown to be:
\begin{equation*}
    \bm{\psi_{\subscript{cMAD}}}(y_i, \bm{x}_i; \bm{\beta}_q, \sigma_q)=
\begin{pmatrix}
    \psi_q\left(\frac{y_i - \bm{x_{i}}^T\bm{\beta}_q}{\sigma_q}\right)\bm{x_{i}} \\
    \text{sign}\left\{\left| \frac{y_i - \bm{x_{i}}^T\bm{\beta}_q - med_i(y_i - \bm{x_{i}}^T\bm{\beta}_q)}{\sigma_q} \right| - \Phi^{-1}(3/4) \right\}
\end{pmatrix}.
\end{equation*}
The cMAD estimator is very similar to the nMAD estimator. However there is one significant difference. The cMAD estimator is less sensitive to changes to $q$. In fact for a null regression model ($\bm{\beta}_q = \beta_{0,q}$) the cMAD estimator is invariant to any variation of $q$ due to the location-shift invariance of the median. In this case the $\beta_{0,q}$ would cancel out of the influence function making the scale estimator only dependent on $y_i$.

Beyond this basic null model the result does not hold, but still it is expected that the cMAD scale estimator will not vary as much across $q$ as the nMAD estimator.

MAD scale estimators are intuitive for symmetric distributions because they correspond to a symmetric interval around the median. However when asymmetry is introduced the MAD estimator loses this intuitive meaning since symmetry is lost. \citet{Rou93} argued that the use of the MAD on highly skewed distributions may be inefficient and impractical. Also in this case the estimator will Huberise more observations on the skewed side of the distribution, which may not be ideal. Further biases arise from the constant $\Phi^{-1}(3/4)$ which ensures the MAD estimator is consistent with the standard deviation when there is a normal distribution. However when the distribution is not normal or when $q \neq 0.5$ this constant loses its relevance. It remains to be seen whether these potential issues actually lead to undesirable properties of the $\bm{\beta}_q$ estimates.

\subsection{Maximum likelihood approach}
M-estimation is just a general form of maximum likelihood estimation, where the loss function is the negative log-likelihood: $\rho(y;\theta) = -\log f(y;\theta)$. So given the known MQ loss function $\rho_q$, which is proportional to the integral of $\psi_q$, the corresponding density function can be calculated using this relationship. \citet{Bia18} explore the parametric distribution associated with the Huber MQ loss function, that plays a similar role to that of the asymmetric Laplace distribution in quantile regression \citep{Yu11,Yu05}. \citet{Bia18} propose the asymmetric least informative (ALI) distribution where there are four parameters: ALI($\mu$, $\sigma_q$, $q$, $c$). Generally $q$ and $c$ is considered fixed, and the density function with the location and scale parameter is:
\begin{equation}\label{ALID}
f(y_i|\mu,\sigma_q)=\frac{1}{\sigma_qB_q}\exp\left\{-\rho_q\left(\frac{y_i-\mu(\bx_i)}{\sigma_q}\right)\right\}, \quad y\in\mathbb{R},
\end{equation}
where
\begin{displaymath}
B_q=\sqrt{\frac{\pi}{q}}\Big[ \Phi(c\sqrt{2q})-1/2 \Big]+\sqrt{\frac{\pi}{1-q}}\Big[ \Phi(c\sqrt{2(1-q)})-1/2 \Big]
\end{displaymath}
 \begin{displaymath}
+\frac{1}{2cq}\exp\{-c^2 q\}+\frac{1}{2c(1-q)}\exp\{-c^2 (1-q)\}.
\end{displaymath}
In the unweighted case ($q=0.5$), the ALI distribution is essentially a modified standard normal distribution with tails (when $|\epsilon_q|>c$) replaced by an exponential distribution. This $q=0.5$ distribution was derived by Huber \citep[][Section 4.5]{Hub81} as the one minimising the Fisher information in the $\varepsilon$-contaminated neighbourhood of the normal distribution. For this reason, it is called the least informative distribution. The properties of the ALI distribution are studied in \citet{Bia18}. The authors also proposed to estimate $\beta_q$, $\sigma_q$ (and $c$) simultaneously by maximising the log-likelihood function:
\begin{equation}\label{loglikeALID}
l_q(y)=-n \log{\sigma_q}-n \log{B_q}-\sum_{i=1}^n \rho_q\left(\frac{y_i-\mu(\bx_i)}{\sigma_q}\right).
\end{equation}
It is also worth noting that $B_q$ is invariant to $\mu$ and $\sigma_q$ so can be removed from the log-likelihood function if $c$ is assumed fixed. In which case the corresponding score vector leads to the influence function:
 \begin{equation*}
    \bm{\psi_{\subscript{ML}}}(y_i, \bm{x_i}; \bm{\beta}_q, \sigma_q)=
\begin{pmatrix}
    \psi_q\left(\frac{y_i - \bm{x_i}^T\bm{\beta}_q}{\sigma_q}\right)\bm{x_i} \\
    \psi_q\left(\frac{y_i - \bm{x_i}^T\bm{\beta}_q}{\sigma_q}\right)\left(\frac{y_i - \bm{x_i}^T\bm{\beta}_q}{\sigma_q}\right) - 1
\end{pmatrix}.
\end{equation*}
When the MQ scale estimator is estimated using this influence function vector we call it the ML estimator. Maximum likelihood estimation has many benefits, however for MQ estimation there are two significant problems. Firstly, it requires a strong distributional assumption which thereby reduces the robustness of the estimator. Having weak distributional assumptions is one of the attractive features of MQ estimation, so it is certainly problematic to add this assumption. Secondly, the second expression in the influence function is not bounded, hence the scale estimator is not outlier-robust. So regardless of the choice of $c$, the ML scale estimator is vulnerable to outliers which contravenes with an attractive property of MQs.

So the ML approach requiring strong distributional assumptions and a non-robust scale estimator makes it an unattractive choice in theory. However, it remains to be seen whether these adverse properties are of much consequence in practice.

\subsection{Method of moments approach}
\citet{Hub64} addressed the problem of estimating a nuisance scale parameter for the Huber estimator. Three proposals were introduced but it was the `Proposal 2' method that was preferred by Huber. The `Proposal 2' method for simultaneously estimating the Huber estimator for the location ($\mu_c$) and scale ($\sigma_c$) parameters requires iteratively solving the following two equations:
\begin{align}
n^{-1}\sum_{i=1}^{n} \psi_{c}\left(\frac{y_i - \mu_c}{\sigma_c}\right) &= 0 \label{eq:p2e1}\\
n^{-1}\sum_{i=1}^{n}  \psi^2_{c}\left(\frac{y_i - \mu_c}{\sigma_c}\right) &= E_{\Phi} \left\{\psi^2_{c}(u)\right\} \nonumber,
\end{align}
which is equivalent to the $M$-estimator with influence function vector:
\begin{equation*}
    \bm{\psi_{c}}(y_i; \mu_c, \sigma_{c})=
\begin{pmatrix}
     \psi_{c}\left(\frac{y_i - \mu_c}{\sigma_c}\right) \\
    \psi^2_{c}\left(\frac{y_i - \mu_c}{\sigma_c}\right) - E_{\Phi} \left\{\psi^2_{c}(u)\right\}
\end{pmatrix},
\end{equation*}
where the $\Phi$ indicates it is the expected value when $u \sim \mathcal{N}(0,1)$. This `Proposal 2' was based on the minimax solution for robust scale estimation where the estimator  $\hat{\sigma}_c$ remains unbiased when $y \sim \mathcal{N}(0,1)$. It can also be viewed as method of moments estimation to ensure that $\hat{\mu}_c$ and $\hat{\sigma}_c$ are both consistent with the parameters of the normal distribution when $y$ is normal.

\citet{Hub77} preferred the `Proposal 2' to the MAD approach in the context of regression because it fit best into the least squares framework and is computationally efficient; sentiments also shared by \citet{Sch80}. \citet[p. 105]{Ham86} stated in reference to \citet[p. 239]{And72} that the MAD approach was shown to be superior, however no such claims were made. \citet{And72} merely stated that three-point descending $M$-estimators are an improvement on `Proposal 2' estimators with large $c$, without any direct comparison to the MAD approach.

Through the use of the ALI distribution the natural extension of the method of moments for MQ regression would be to use the following influence function vector defining the MQ method of moments (MM) estimator:
 \begin{equation}\label{eq:infMM}
    \bm{\psi_{\subscript{MM}}}(y_i, \bm{x_{i}}; \bm{\beta}_q, \sigma_q)=
\begin{pmatrix}
    \psi_q\left(\frac{y_i - \bm{x_{i}}^T\bm{\beta}_q}{\sigma_q}\right)\bm{x_{i}} \\
    \psi_q^2\left(\frac{y_i - \bm{x_{i}}^T\bm{\beta}_q}{\sigma_q}\right) - E_{\Omega} \left\{\psi^2_q(u)\right\}
\end{pmatrix},
\end{equation}
where the $\Omega$ indicates it is the expected value when $u \sim ALI(0, 1, q, c)$. By making this distributional assumption it ensures that the estimators are consistent with the first two moments of the ALI distribution, i.e. the location and scale parameter. Unlike the ML approach the distributional assumption is only required for the first two moments rather than the full distribution. The standardisation property of the ALI distribution, i.e. if $u \sim ALI(\mu, \sigma, q, c)$, then $(u - \mu)/\sigma \sim ALI(0, 1, q, c)$, ensures consistency of the estimates.

To simultaneously get estimates of $\bm{\beta}_q$ and $\sigma_q$ for the MM approach, iterations for both estimates must be incorporated into the IRLS algorithm. Let $\hat{\bm{\beta}}_q^{[m]}$ and $\hat{\sigma}_{q}^{[m]}$ be trial values for $\hat{\bm{\beta}}_q$ and $\hat{\sigma}_q$. For the MM approach the scale iteration will be:
\begin{equation}
  \left(\hat{\sigma}_{q}^{[m+1]}\right)^2 = \frac{1}{(n-p)E_{\Omega} \left\{\psi^2_q(u)\right\}}\sum_{i=1}^n \psi_q\left(\frac{y_i - \bm{x}_i^T\hat{\bm{\beta}}_q^{[m]}}{\hat{\sigma}_{q}^{[m]}}\right)^2 \left(\hat{\sigma}_{q}^{[m]}\right)^2
\end{equation}
where eventually the iterations will converge on a solution.

\section{Assessment of scale estimators}\label{sec:scalecomp}
The proposed MM approach provides an alternative to the pre-existing nMAD, cMAD and ML approaches to MQ scale estimation. It remains to be seen which of the four approaches performs the best in practice. The quality of an estimator is typically assessed by which has the minimum mean squared error (MSE). All four approaches have estimating equations which ensure that the location MQ estimates are unbiased, so provided the algorithm converges it can be assured there is no bias. Therefore to ascertain which of the four approaches has the minimum MSE only the variance needs to be considered. However, assessing the performance of these approaches solely on which estimator has the least variance would be a very narrow assessment. For MQs, we suggest two other key properties of the estimator that need to be assessed:
\begin{itemize}
\item How smoothly the MQ estimates transition from the quantile to the expectile estimates as $c$ increases.
\item Whether outlying observations are Huberised appropriately across all values of $q$.
\end{itemize}
Furthermore, since MQs do not have strong distributional assumptions it is desirable that these properties are maintained across a range of distribution types. To assess how well the four approaches exhibit these properties a simulation study is performed.

\subsection{Description of the simulation}
For simplicity the simulations were performed on null models where the only regression coefficient to be estimated is $\beta_{0,q}$, i.e. the intercept with no covariates. Although this is not realistic in practice, the effect of different scale estimators will be most prominent on the intercept coefficient, hence slope coefficients will only unnecessarily complicate the simulation study. Also, a linear, homoscedastic model would yield the same slope coefficients regardless of scale estimator. 

The four MQ approaches are tested on three different data distributions: the standard normal, standard log-normal (highly skewed) and the t-distribution with $df = 3$ (heavy-tailed). Three different tuning constants are used, $c= 0.5, 1.3, 3.0$, to capture any differences caused by $c$. And finally a uniform array of values for $q=0.01, \dots, 0.99$ are used for all simulations. In the simulation, estimates of $\beta_{0,q}$ and $\sigma_{q}$ are obtained by taking the mean of 100 simulated estimates each from a sample size of $n=10^4$. Based on these simulated estimates four distinct properties of the MQ estimators are assessed:
\begin{enumerate}
\item The general effect of the scale estimation method on the location $\beta_{0,q}$ and scale $\sigma_{q}$ estimates across the distributions, $q$ and $c$. 
\item The asymptotic variance is estimated for each of the three distributions and each chosen $q$ and $c$. Note that this is estimated using plug-in estimates to a sandwich estimator from a single simulated sample of $n = 10^6$, not the simulated estimates. The derivation of the asymptotic variance can be requested from the authors.
\item The smoothness of the transition from the quantile to the expectile estimates as $c$ increases.
\item The proportion of Huberised residuals across $q$ and $c$.
\end{enumerate}
Generally $n$ is chosen to be rather large to ensure that the true trends are captured. And then only 100 replicates are performed as any more offer no added stability. Further descriptions and results of these simulations are reported below.

\subsection{M-quantile estimates}
\begin{figure}
\centering
\includegraphics[width = \textwidth]{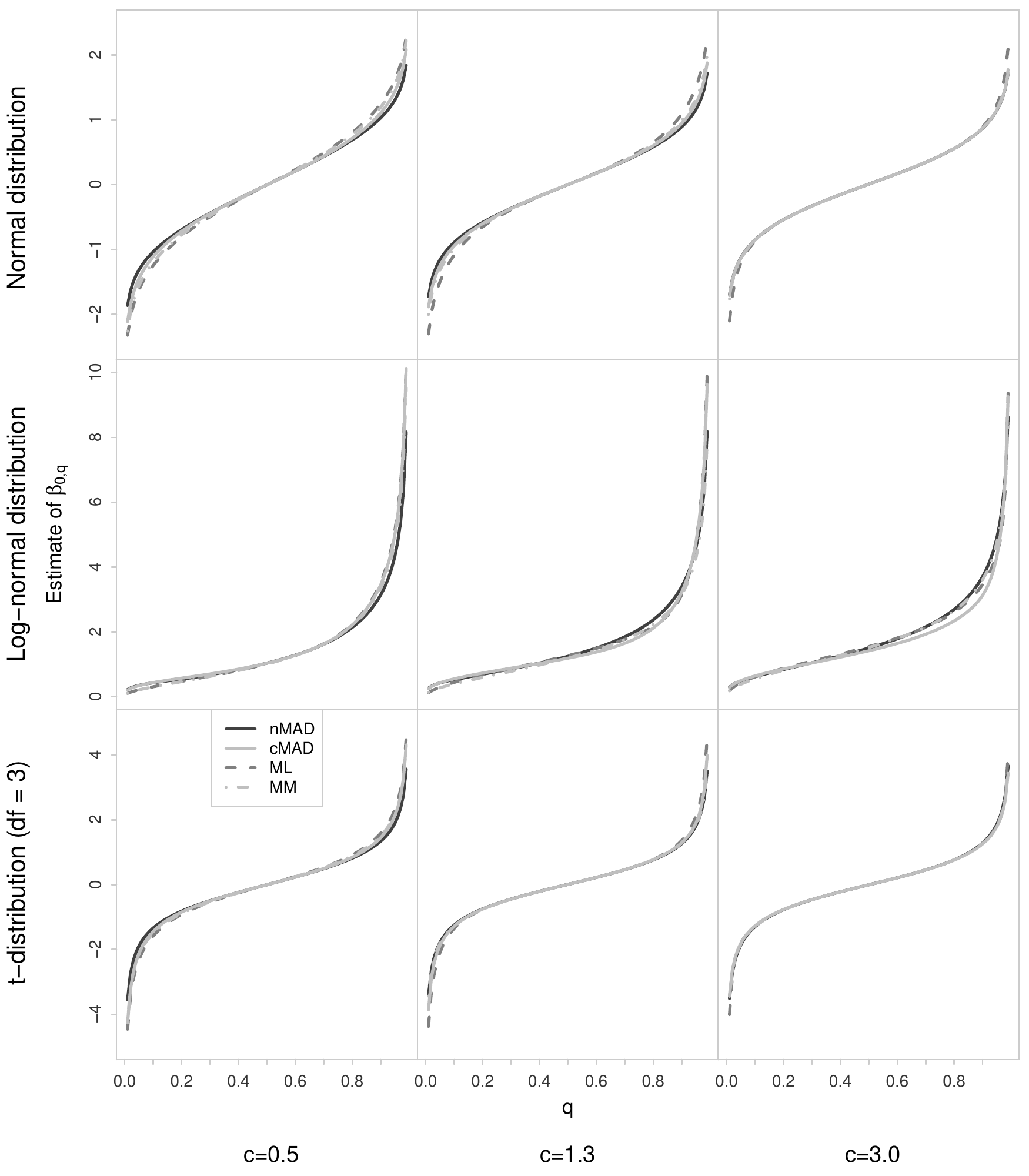}
\caption{Estimates of the location parameter $\beta_{0,q}$ across $q$ for three different distributions and $c=0.5, 1.3, 3.0$.}
\label{fig:locations}
\end{figure}

\begin{figure}
\centering
\includegraphics[width = \textwidth]{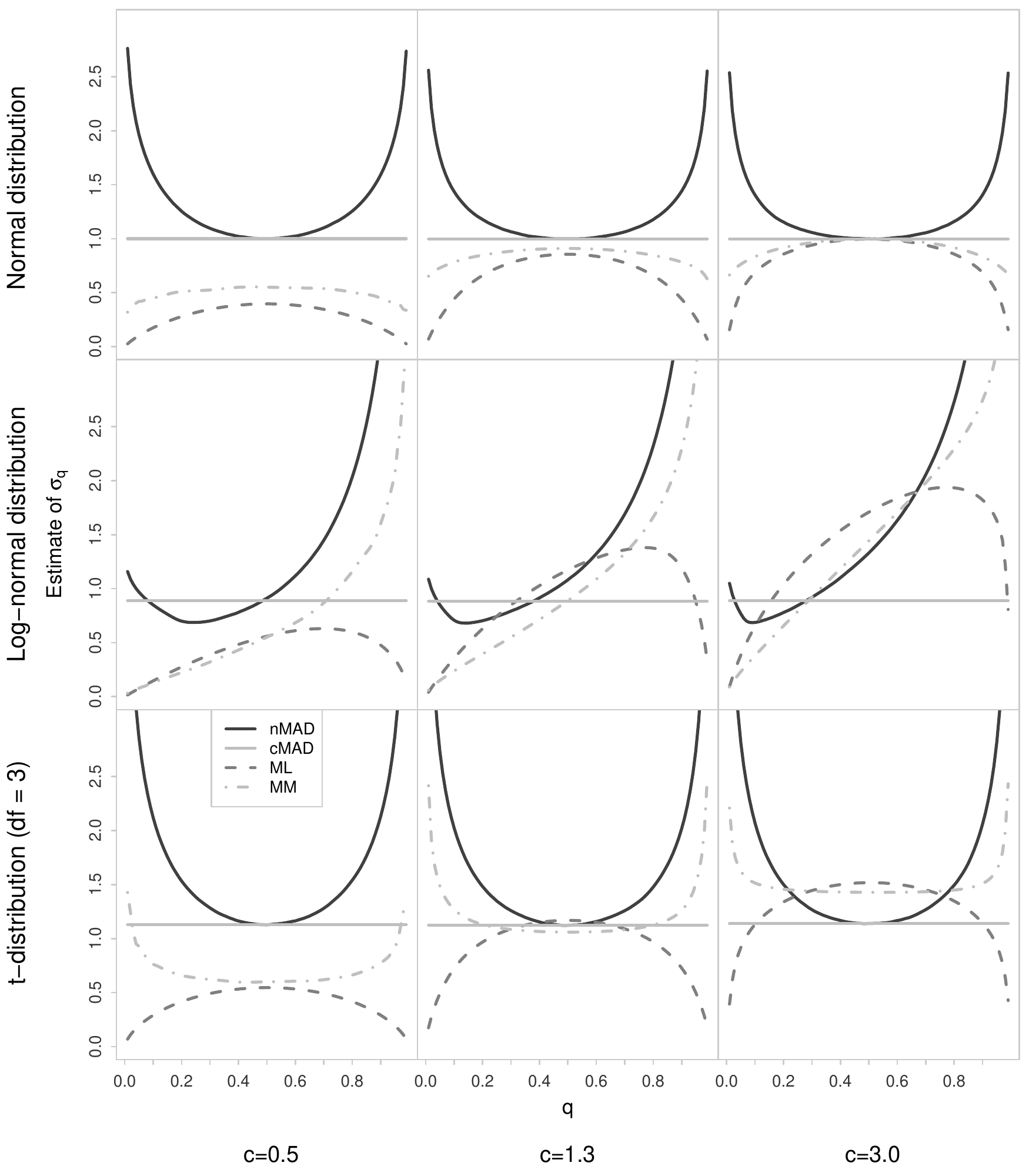}
\caption{Estimates of the scale parameter $\sigma_{q}$ across $q$ for three different distributions and $c=0.5, 1.3, 3.0$.}
\label{fig:scales}
\end{figure}

The location $\beta_{0,q}$ and scale estimates $\sigma_{q}$ for the three different distributions, $q$ and $c$ are shown in Figures~\ref{fig:locations} and \ref{fig:scales} respectively.

From what can be observed in Figure~\ref{fig:locations} the location estimates are relatively similar across all four approaches. A closer inspection reveals that the nMAD estimates do not vary as much across $q$, compared to the ML approach which varies the most. For the normal and t-distribution the estimates of the four approaches converge closer together as $c$ increases, whereas the opposite appears to occur for the log-normal distribution.

Conversely to location estimates, there appears to be substantial differences in the scale estimates as shown in Figure~\ref{fig:scales}.  As already stated, the scale estimates for the cMAD approach are invariant to changes in $q$ and $c$, as is displayed in the figure. However the other three approaches change quite substantially across $q$ and $c$. The nMAD approach generally estimates the largest scale estimates, especially as $q$ becomes extreme where the scale estimates become increasingly large. Hence the nMAD scale estimates have this convex shape for all distributions across $q$. For symmetric distributions the nMAD and cMAD approaches have the same scale estimates when $q=0.5$. The ML approach generally estimates the lowest scale values, and as $q$ becomes extreme it gets closer to 0 which gives a concave shape for all distributions across $q$. Generally, the MM estimates lie in between the nMAD and ML estimates. Interestingly, the scale estimates for the MM approach differ in shape across $q$, with a concave shape for the normal distribution, but a convex shape for the other distributions. The convex and concave shapes occur due to differences in the actual data distribution compared to the distribution in which the estimator is designed to be consistent with. For extreme $q$ this disparity is at its greatest. Lastly, as $c$ increases, the scale estimates for the ML and MM approach also increase, while the nMAD appears almost invariant to changes in $c$. 

\subsection{Asymptotic variance}
\begin{figure}
\centering
\includegraphics[width = \textwidth]{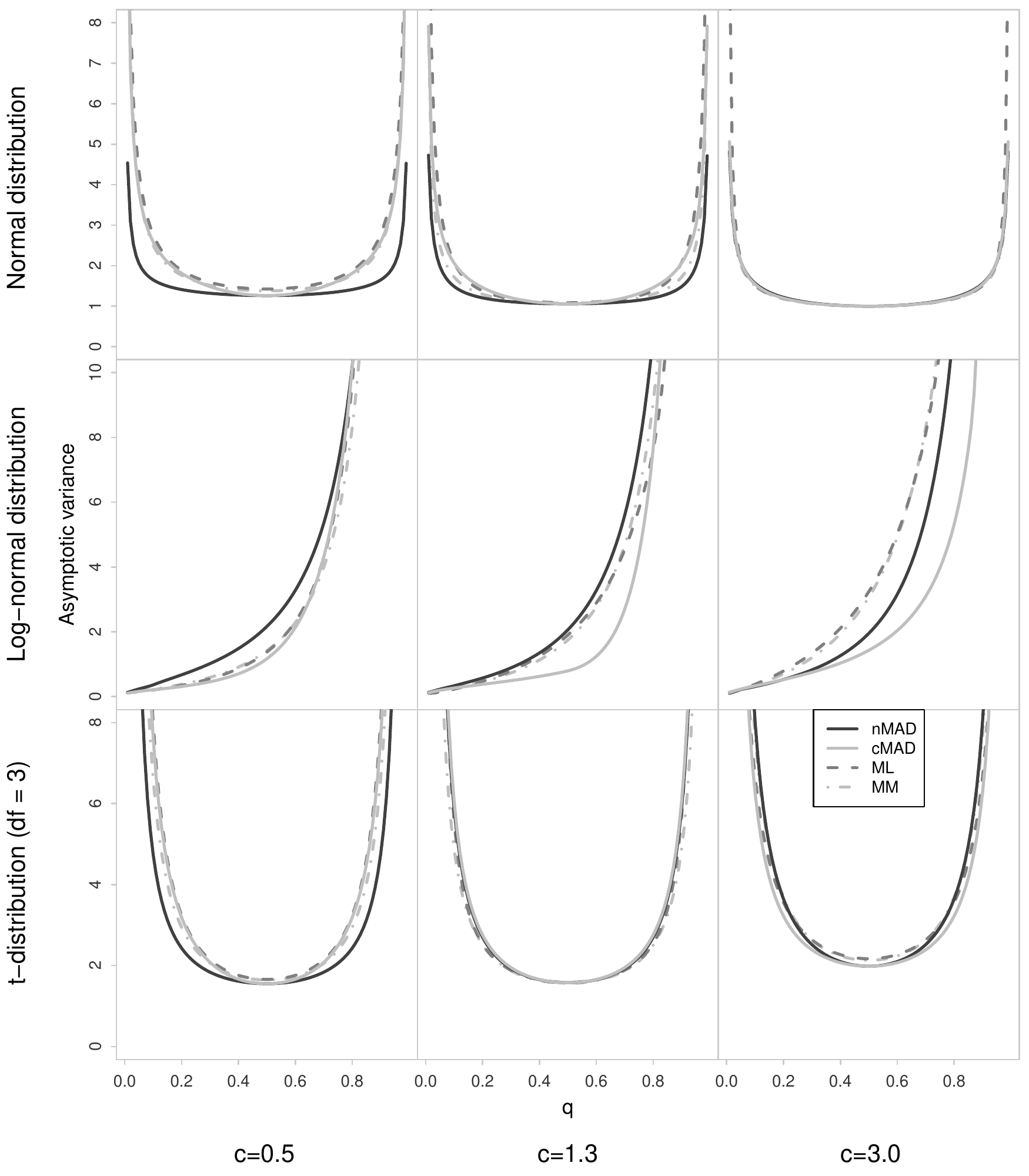}
\caption{Asymptotic variances of the MQ intercept estimator $\hat{\beta}_{0,q}$ for each of the four approaches for three different distributions, three different tunings constants and across values of $q$.}
\label{fig:varmu}
\end{figure}

The asymptotic variance of the MQ estimates will depend not only on the data distribution but also the values of $q$ and $c$. The asymptotic variance of the location $\beta_{0,q}$ for each of the four approaches is shown in Figure~\ref{fig:varmu}.

Generally the asymptotic variances of $\beta_{0,q}$ are reasonably similar across the four approaches, especially for the t-distribution. For the normal distribution the smallest variance is from the nMAD approach, particularly with smaller $c$. The other three approaches have only minor differences for the normal distribution across $c$. The log-normal distribution reveals the cMAD approach to generally have the smallest variance, which becomes more pronounced as $c$ increases. The ML and MM approaches yield almost identical variances across all distributions and variations of $c$. But overall, the MAD approaches generally have a lower asymptotic variance for $\beta_{0,q}$ compared to the ML and MM approaches which are very similar.

\subsection{Smooth transition from quantile to expectile}\label{sec:q2e}
As mentioned, the Huber MQ estimator provides a versatile middle-ground between the quantile where $c\rightarrow 0$, and the expectile where $c\rightarrow \infty$. Hence a favourable property of MQ estimators is that the estimates transition smoothly from the quantile to the expectile as $c$ increases.

\begin{figure}
\centering
\includegraphics[width = \textwidth]{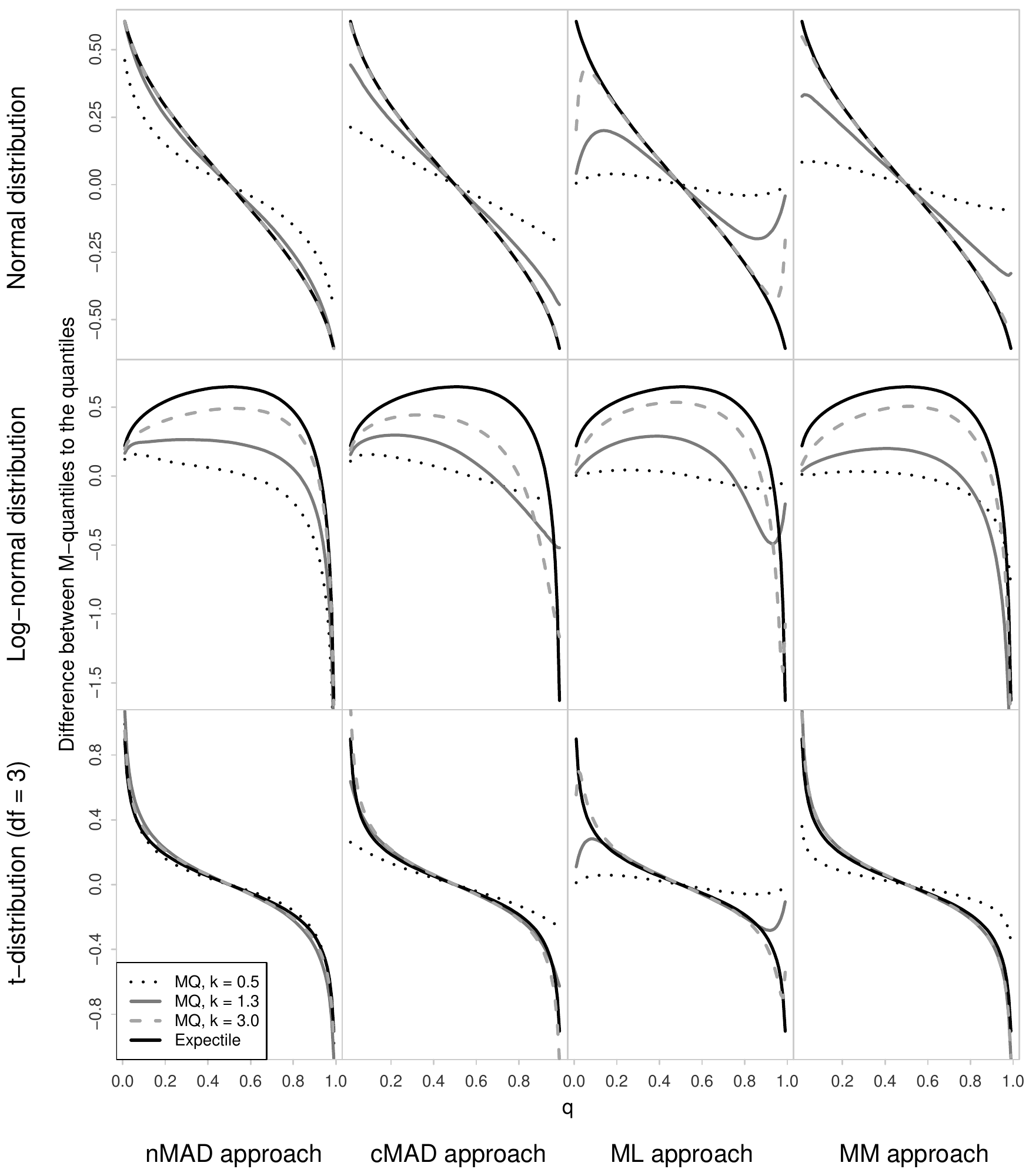}
\caption{Expectile and $\beta_{0,q}$ estimates for $c = 0.5, 1.3, 3.0$ compared to the quantile (at $y=0$).}
\label{fig:distk}
\end{figure}

Figure~\ref{fig:distk} demonstrates how smooth this transition is by comparing the estimates of the MQ intercept parameter $\beta_{0,q}$ of the four approaches to the scale-invariant quantiles and expectiles as $c$ increases. The $y$-axis in the figure shows the difference between the MQs and the quantiles, hence $y=0$ represents the quantile estimates. An ideal $\beta_{0,q}$ will be close to the quantile when $c=0.5$ and close to the expectile at $c=3.0$ with the estimate for $c=1.3$ in between.

For all three distributions with $c=0.5$ the nMAD approach returns $\beta_{0,q}$ estimates which are the furthest from the quantile. Furthermore, estimates for the nMAD approach on the t-distribution are all very similar across all three values of $c$. This suggests that the nMAD approach is the least sensitive to variations in $c$.

There are clear problems with the ML approach when $q$ is close to 0 or 1. Even when $c=3.0$, the $\beta_{0,q}$ estimates for extreme $q$ tend towards the quantile. This is especially noticeable in the normal distribution. The ML approach clearly does not transit uniformly from quantile to expectile as $c$ increases. This is due to the very small estimates of $\sigma_{q}$ which occur for extreme $q$. Very small values of $\sigma_{q}$ result in high proportions of observations being Huberised (which is further explored in the next section), and the more observations that are Huberised by $c$ the closer the estimate will be to the quantile. This constraining feature reveals that the ML approach is not appropriate with extreme $q$, such as when $q < 0.2$ or $q > 0.8$.

The cMAD and MM approach perform similarly and both quite well. They each show a suitable level of sensitivity to $c$ and transition smoothly from quantile to expectile across $q$. For all three distributions, estimates for $c=0.5$ are close to the quantile, and for $c=3.0$ are close to the expectile. Even the log-normal distribution with its extreme skewness still maintains an appropriate looking transition for the cMAD and MM approaches.

\subsection{Proportions of Huberised residuals}\label{sec:propwin}
Traditional robust methods aim to reduce the influence of heavy-tailed outliers when estimating the centre of a distribution. However, when estimating MQs with extreme $q$ the estimate is already near the tail, and it would be unreasonable to reduce the influence of observations close to the actual MQ estimate. The tuning constant should be set to Huberise just the outliers, and not too many of the observations closer to the estimate.

\begin{figure}
\centering
\includegraphics[width = \textwidth]{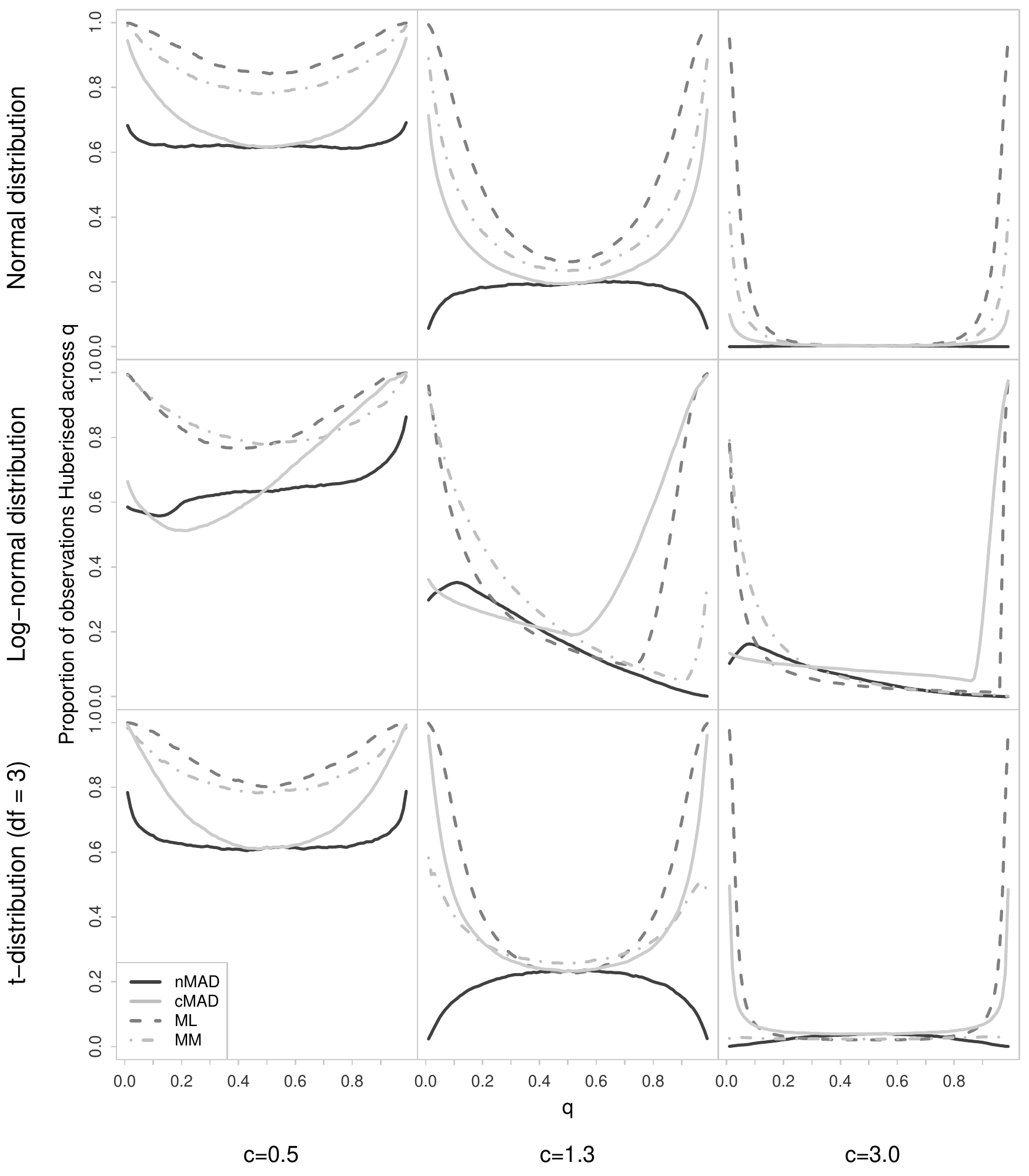}
\caption{Proportion of Huberised residuals across $q$ for three different distributions and $c= 0.5, 1.3, 3.0$. }
\label{fig:propwin}
\end{figure}

The proportion of Huberised observations across $q$ is an important indicator for determining the appropriateness of an MQ estimator. An appropriate MQ estimator should Huberise observations somewhat similarly across $q$, in order to ensure that $c$ should have a similar effect across all $q$. For example, if at $q=0.5$, 60\% of observations are Huberised but at $q=0.95$ almost no observations are Huberised then this may be problematic should some robustness be required for all $q$. So the more dynamically that observations are Huberised across $q$, the less suitable a global $c$ for all $q$ will be. Figure~\ref{fig:propwin} reveals the proportions of Huberised observations for each of the four approaches.

The general trend for all approaches is that fewer observations are Huberised as $c$ increases which is expected, and also fewer when $q$ is closer to 0.5.

The most variability in proportions across $q$ is found in the ML approach, especially when $c=3.0$. For the ML approach the proportion of Huberised observations is always close to 1 at extreme $q$, regardless of $c$. This is an undesirable property. Although when $c=0.5$ the ML approach Huberises more evenly across $q$, it generally is problematic in that it Huberises all observations as $q$ gets closer to 0 or 1.

In general the nMAD approach Huberises the observations less than the other three approaches. And importantly, this approach has proportions which vary the least across $q$. The one concerning aspect of the nMAD approach is that for $c=1.3$ very few observations are Huberised at extreme values of $q$. This means that for a given $c$, the MQs for the nMAD approach are not as robust to outliers compared to the other approaches for extreme $q$. This explains why the nMAD MQ estimates remained quite similar to the expectile in the previous section.

The cMAD and MM approaches generally Huberise the observations in a convex shape across $q$, with the cMAD approach being more variable. The log-normal distribution for $c=3.0$ shows that for the cMAD approach, a high proportion of observations are Huberised for $q$ close to 1. For the log-normal and t-distribution the MM approach has the least variation across $q$ and varies relatively similarly to the other approaches for the normal distribution. Hence in general it suggests that the MM approach Huberises observations the least erratically across different distributions and values of $q$.

\subsection{Summary of M-quantile scale estimators}
So although the MQ estimates in Figure~\ref{fig:locations} appeared all quite similar across all four approaches, there are considerable differences caused by the different scale estimators. These differences arise in other properties which were explored in the simulation study. The nMAD and ML approaches were revealed to have significant deficiencies as MQ scale estimators. For the nMAD approach the problem was a lack of sensitivity to different values of $c$, while the ML approach is not robust and for extreme $q$ too many observations are Huberised. However, the cMAD and the proposed MM approach were revealed to perform relatively well. The cMAD approach is a simple approach which offers invariant scale estimates across $q$ and $c$ which is a favourable property considering the interrelationship between the two in how they Huberise. The MM approach generally performs rather similarly to the cMAD approach but it lacks the simplicity.

While all approaches offer some useful attributes, generally the cMAD and MM approach appear to be the best MQ approaches in the simulation study. Hence either of these approaches should be used for MQ estimation. As it is ideal to have a single approach to be used as a default choice, the authors recommend cMAD over the MM approach. This is solely due to the relative conceptual simplicity of the cMAD approach, and in practice the estimates from the two approaches are so similar anyway. 

\section{Tuning constant selection methods} 
In the past, Huber MQ estimation has relied on users to select the tuning constants based on their own subjective choice. In this section two methods are introduced which aim to select an optimal adaptive or data-driven tuning constant based on pre-defined criteria. There are two types of tuning constant selection methods: a local and a global method. Local tuning constant selection finds a tuning constant for a specific $q$, whereas global selection requires fitting an ensemble of MQ models across a grid of many values of $q$, and finding a single tuning constant for the entire ensemble. Such an ensemble modelling technique is used when using MQ models in SAE.

The first of the two methods proposed is a local method based on selecting a tuning constant that optimises the efficiency of the estimates. The second is a global method based on reducing contamination using an inverse MQ function. The benefit of these methods is that the tuning constant can be selected without subjective user choice, hence making MQ regression more user-friendly and also leading to more efficient estimation. It is worth mentioning that there is a third method introduced by \citet{Bia18} which is based on selecting the tuning constant which maximises the log-likelihood function, however it was shown in the previous section that maximum likelihood was not appropriate for MQ estimation hence we do not address this method.

Not only does a tuning constant selection method prevent subjective choices, it also provides a useful diagnostic informing on the level of contamination in the data. If the selected tuning constant is large (e.g. $c > 3$) then this indicates that contamination is low, while the closer to zero it is the higher the level of contamination. This provides another motivation to introduce these methods. 

The two adaptive tuning constant selection methods for MQ regression are described in more detail below.

\subsection{A local method for M-quantile tuning constant selection}\label{sec:3}
\citet{Wan07} proposed a method for tuning constant selection for the Huber estimator, which is a special case of the Huber MQ with $q=0.5$. It is therefore simple to extend these methods for a general $q \in (0,1)$. \citet{Wan07} proposed an efficiency factor $\tau$ for the Huber estimator as defined by the influence function in Equation~\eqref{eq:psi_huber}:
\begin{equation} \label{eq:eff}
\tau = \frac{E[\psi_c'(\epsilon)]^2}{E[\psi_c^2(\epsilon)]},
\end{equation}
where $\epsilon$ are the residuals. This is an efficiency factor because it has an inverse relationship with the variance of the regression parameters. Hence the maximum value of $\tau$ for a given data set yields the least variance of the regression estimator. An MQ efficiency factor can be generalised to MQs, by replacing $\psi_c$ with $\psi_q$:
\begin{align*}
\tau_q &= \frac{E[\psi_q'(\epsilon_q)]^2}{E[\psi_q^2(\epsilon_q)]} \\
&= \frac{[\int_{-c}^c\psi_q'(\epsilon_q)dF(\epsilon_q)]^2}{\int_{-\infty}^{\infty}\psi_q^2(\epsilon_q)dF(\epsilon_q)}.
\end{align*}
Using the law of large numbers the estimator of $\tau_q$ can then be derived:
\begin{align*}
\hat{\tau}_q &=\frac{n^{-2}\sum_{i=1}^n[\psi_q'(\hat{\epsilon}_{iq})I(-c \leq \hat{\epsilon}_{iq} \leq c)]^2}{n^{-1}\sum_{i=1}^n\psi_q^2(\hat{\epsilon}_{iq})}\\
&=\frac{4n^{-2}\sum_{i=1}^n (1-q)^2I(-c \leq \hat{\epsilon}_{iq} < 0) + q^2I(0 \leq \hat{\epsilon}_{iq} \leq c)}{n^{-1}\sum_{i=1}^n\psi_q^2(\hat{\epsilon}_{iq})I(-c \leq \hat{\epsilon}_{iq} \leq c) + 4(1-q)^2c^2I(\hat{\epsilon}_{iq} < -c)+ 4q^2c^2I(\hat{\epsilon}_{iq} > c)}.
\end{align*}

With an efficiency factor estimator derived the tuning constant which maximises $\hat{\tau}_q$, and hence minimises $Var(\hat{\bbeta}_q)$, can then be found using these steps:
\begin{enumerate}
\item Obtain initial estimates $\hat{\bbeta}_q, \hat{\sigma}_q$ using the cMAD MQ regression model and an initial value of $c$, such as $c=1.3$. Note that initial values might converge on a local maximum when $q$ is extreme and $c$ is either very small or very large. Hence a modest starting point of 1.3 is advised.
\item Calculate the $q$-th MQ residuals with $\hat\epsilon_{iq}=(y_i-\bm{x}_i^T\hat{\bbeta}_q)/\hat{\sigma}_q.$
\item Compute $\hat{\tau}_q(c)$ over a grid $g=[0.5,4]$ of $c$ values. Anything less than 0.5 becomes unstable.
\item Select the $c$ value that maximizes $\hat{\tau}_q$.
\item Repeat steps 1-4, except using the new value of $c$ instead of the initial choice.
\item Stop once the value of $c$ that maximizes $\hat{\tau}_q$ remains unchanged.
\end{enumerate}
The selected tuning constant for this method ensures that the efficiency of the regression estimators is maximised for any given choice of $q$.

\subsection{A global method for M-quantile tuning constant selection}\label{sec:invMQ}
The functional form of an MQ estimator, $Q_{q}$, based on Equation~\eqref{eq:esteq}, for a given distribution function $F(\cdot)$ is defined as:
\begin{equation}\label{eq:MQfunct}
\int{\psi_q\left(\frac{Z-Q_q}{\sigma_q} \right)F(dz)}=0.
\end{equation}
\citet{Jon94} showed that expectiles, and more generally MQs, are themselves quantiles of not $F(\cdot)$, but a different distribution $G(\cdot)$. In other words, for a given MQ with known influence function and known $F(\cdot)$, $G(Q_q) = q$. Hence $G(\cdot)$ can be considered to be an inverse MQ function where instead of choosing $q$ and deriving $Q_{q}$ as is typical, one can choose $Q_{q}$ and derive $q$.

\citet{Jon94} proves that this function $G(\cdot)$ is actually itself a distribution function, and shows that $G(\cdot)$ can be derived by expanding the integral in Equation~\eqref{eq:MQfunct} and rearranging to make $q$ the subject. For the Huber MQ with a given $c$ this results in:
\begin{equation} \label{eq:G}
 G_c(x;F) = \frac{ b^- F_y(a^-) + \frac{1}{\sigma_q}\left[H(x)-H(a^-)- x F_y(x)\right]}  {\frac{1}{\sigma_q}\left[2H(x)-H(a^-)-H(a^+)- 2x F_y(x)\right] + b^- F_y(a^-) +  b^+ F_y(a^+) - c},
\end{equation}
where $a^- = x-c\sigma_q$, $a^+ = x+c\sigma_q$, $b^-=\frac{x}{\sigma_q}-c$, $b^+=\frac{x}{\sigma_q}+c$, and $H(t) = \int_{-\infty}^{t} y dF(y)$. A complete derivation is shown in Appendix~\ref{sec:appendix}. Hence under a normality assumption, with $F_N(\cdot) \sim N(\mu, \sigma)$, then $G_c(\cdot;F_N)$ is well defined since $H(t;F_N)=\mu \Phi \left(\frac{t-\mu}{\sigma} \right) - \sigma \phi \left(\frac{t-\mu}{\sigma} \right)$. In this case, if $\hat{Q}_{q}$ is the MQ estimate from a normal random variable then $G_c(\hat{Q}_{q}; F_N) = q$. This is an important point for selecting a tuning constant when data is contaminated.

Consider the case of a contaminated normal distribution $F_{cont}\sim (1-\alpha)\Phi + \alpha U$, where $U$ is an unknown contaminating distribution and $0 \leq \alpha < 1$ is a known level of contamination. Robust estimation aims to limit the effect of the contaminated distribution. With Huber MQ estimation this contamination is limited using the tuning constant $c$. Clearly, the higher the contamination level $\alpha$, the lower $c$ will need to be. Ideally, the robust estimates on $F_{cont}$ aim to be as close to $\Phi$ as possible. In other words, the contaminated distribution should be down-weighted, and similarly $c$ selected, in such a way that the contamination effect is negligible.

As stated, the inverse MQ function can verify whether the MQ estimates are approximately normal or not. If there is contamination then these estimates are at risk of being affected by the contamination. However, with an appropriate tuning constant $c$, the estimates will not be as susceptible to the contamination. So in the contaminated normal case, the optimal $c$ will be such that the MQ estimates of $F_{cont}$ are as close as possible to $\Phi$. Hence finding this optimal $c$ can be found using the following steps:
\begin{enumerate}
\item Define a grid of $c$ values (e.g., $[0.5,4]$). and let $c_j$ be the $j$-th value of $c$ in the grid.
\item For each $c_j$, calculate MQ estimates $\hat{Q}_{q}(\mathbf{x}_{i})$ with the cMAD scale estimator $\hat{\sigma}_{q}$ for a uniform grid of $q$, e.g. $q=0.01, 0.02, \dots, 0.99$.
\item For each $c_j$, calculate:
\begin{equation*}
\hat{q}^+=G_{c=4}\left( \hat{\sigma}_{q}^{-1}\left[med_i\left\{\hat{Q}_{q,c_j}(\mathbf{x}_{i})\right\} - med_i\left\{\hat{Q}_{q=0.5,c_j}( \mathbf{x}_{i})\right\}\right];\Phi \right).
\end{equation*} 
This calculates the inverse MQ function assuming normal residuals, and hence an arbitrarily high tuning constant of $c=4$ is set as this is appropriate with normal data. 
\item For each $c_j$ calculate $d_j=\sum_q\left(\hat{q}^+ - q\right)^2$ as a measure of the deviation of $\hat{q}^+$ from $q$. The smaller $d_j$ is, the less deviation. The absolute value of the deviation could also be used, though this is not likely to affect the results since large deviations are unlikely.
\item Find $j_{min}$ such that $d_{j_{min}}$ is the minimum value of $d_j \forall j$.
\item Choose $c_{j_{min}}$ as the optimal tuning constant.
\end{enumerate}
So the inverse MQ function $G_{c=4}(\cdot;\Phi)$ in step 3 is used as this would be appropriate if the data were normal. If this were the case then the MQ estimates with arbitrarily large $c=4$ would result in no deviation between $\hat{q}^+$ and $q$. However the contaminating data ensures there is some deviation. This method finds the value of $c$ which corresponds with MQ estimates being the closest to the theoretical MQ estimates under no contamination. The closer that $\hat{q}^+$ is to $q$, the closer the MQ estimates of the contaminated data are to what they would be under no contamination. So this method can be viewed as finding $c$ which minimises the effect of the contamination on the MQ estimates.

\section{Simulation studies}\label{sec:sim}
In this section we assess the performance of the data-driven tuning constant methods at $q=0.25, 0.5$ and 0.75 based on a Monte-Carlo simulation study. The objective of this simulation study is threefold. First, we investigate the ability of the proposed data-driven tuning constant methods to account for different levels of contamination in the data. Second, we assess the influence of different samples sizes on the proposed approaches. Third, we compare the efficiency of the MQ estimates using data-driven and fixed tuning constants. For all aims, data are generated under an independent linear model
$$y_i=\mathbf{x}^T_{i}\bbeta +\epsilon_i, \enspace\enspace i=1,...,n,$$
where $\bbeta = (\beta_0, \beta_1)^T=(100,4)^T$ and the single covariate follows a normal distribution with $\mu_x=1$ and $\sigma_x=1$. The sample sizes are set to $n=100, 500, 1000$ and $10,000$. Three different settings for the error distribution $\epsilon_i$ are considered:
\begin{itemize}
\item Contaminated normal distribution: For normally distributed errors, we investigate different levels of contamination generated by
$$\epsilon_i\sim (1-\alpha) N(0,1) + \alpha N(0,150),$$ where $\alpha$ is set to $0\%, 5\%$ and $20\%$.
\item $t$-distribution: $\epsilon_i\sim t(3)$.
\item Cauchy distribution: $\epsilon_i\sim \text{Cauchy}(\text{location}=0, \text{scale}=1)$.
\end{itemize} Under scenario $\text{Normal}\,0\%$ the assumption of normality is obviously valid, whereas the other settings define situations with clear departures from normality to more heavy-tailed distributions. Each setting is repeated independently $500$ times. We derive MQ estimates using data-driven tuning constants based on the local asymptotic variance method (MQ AV) and the global inverse MQ approach (MQ Inv) introduced in Sections \ref{sec:3} and \ref{sec:invMQ} respectively. For the third aim of the simulation study, fixed tuning constants of 1.345 and 4 were used for comparisons of efficiency. The usual default choice of $c=1.345$ serves as a benchmark.

\begin{figure}[ht]
\begin{centering}
\includegraphics[width=1 \textwidth]{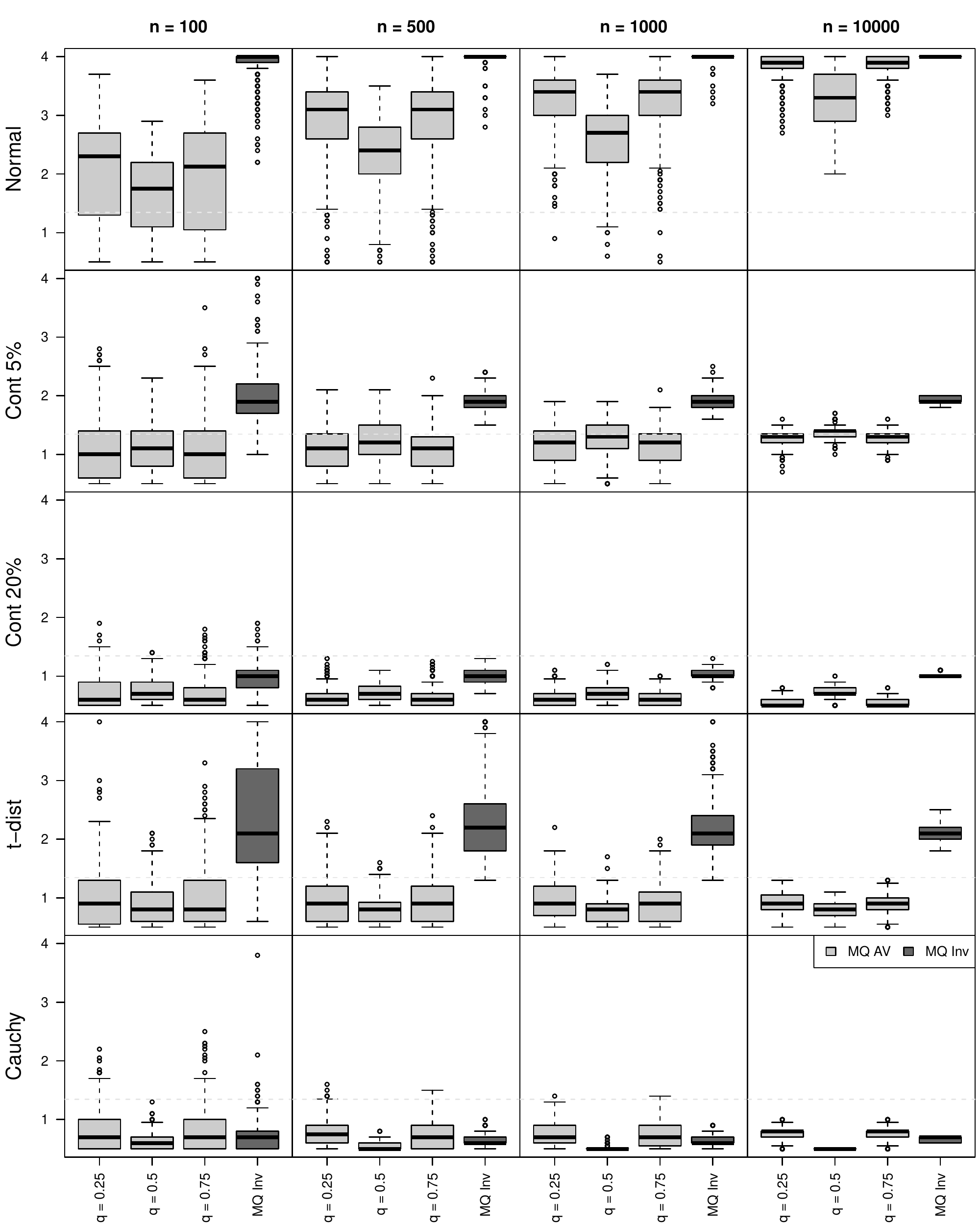}
\caption{\label{TunConst_n10000}Tuning constants under the different distributions and sample sizes.}
\end{centering}
\end{figure}

Starting with the first and second aim, we investigate the behaviour of the proposed data-driven tuning constants across the different distributions and sample sizes. Figure \ref{TunConst_n10000} shows the distribution of the estimated tuning constants based on the MQ AV and MQ Inv methods in the different scenarios and sample sizes at $q=0.25, 0.5$ and 0.75 over the 500 replications. The horizontal line represents the usual choice of $c=1.345$. The light-grey boxes indicate the optimal $c$ based on the MQ AV approach, whereas the dark boxes show the values of $c$ for the MQ Inv method. A reminder that the MQ Inv approach selects a global tuning constant for the entire distribution, hence is not reported for specific values of $q$. 

Under the scenario with an uncontaminated normal distribution, we observe that the MQ AV and MQ Inv lead to tuning constants clearly larger than 1.345 for all $q$. This is expected because under this scenario the assumptions of the linear model hold. Therefore, there is no required resistance against outliers and so a finite tuning constant is not required. In contrast, for the contaminated normal distributions the tuning constants of the MQ AV and the MQ Inv are smaller, often less than $c=1.345$. Both methods are able to select tuning constants which reflect different levels of contamination in the data. For instance, the median optimal tuning constant for the MQ AV at $q=0.5$ and $n=10,000$ reduces from 1.4 to 0.7 when the level of contamination increases from $5\%$ to $20\%$. This is consistent with what we expect because more influential observations should be down-weighted as the level of contamination becomes higher. The heavy-tailed Cauchy distribution shows similarly small optimal tuning constants for both methods. For the MQ AV approach there are noticeable differences between $q=0.5$ and $q=0.25, 0.75$, with the differences depending on the distribution. 

As the MQ AV method is based on asymptotic results, the second aim investigates the influence of different sample sizes on the data-driven tuning constants. Ideally the results should not be too affected by sample size. This appears to be the case except for the obvious exception of the uncontaminated normal where the tuning constant results drift to a higher value as $n$ increases. All other distributions result in tuning constants which remain rather similar across all sample sizes. In regard to the variation in the tuning constant, it can be observed that the tuning constants from both methods become more stable when the sample size increases. The tuning constants are generally more stable for $q=0.5$ for the MQ AV method, compared to $q=0.25, 0.75$.

Comparing the two methods we find that the MQ AV method generally results in smaller tuning constants than the MQ Inv method. The one exception is the Cauchy distribution, where they all are confined to the lower bound of 0.5. Also, generally the MQ Inv method has less variation of values of $c$, with a noticeable exception for the t-distribution. This may be an indication of it performing non-ideally due to the assumption of a contaminated normal which underpins the MQ Inv method. 

Having assessed the behaviour of the data-driven tuning constants, the third aim of this simulation study is to evaluate if the MQ estimates based on data-driven tuning constants lead to more efficient results. To do this the efficiency of the coefficient estimates are presented in two ways: the simulated variation and the analytic variation. The simulated variation is measured by the MAD of the 500 simulated coefficient estimates, the MAD is chosen rather than the standard deviation because the coefficient distribution was heavy-tailed. The analytic variation is measured by the median of the asymptotic standard error estimates based on the inverse function in Equation~\eqref{eq:eff}. Both these measures of variation were assessed for the estimates based on the optimal data-driven tuning constants as well as pre-selected $c=1.345, 4.0$. We use two measures to improve reliability of the estimates, it also provides a comparison between the analytic standard error and the simulated standard error which can verify the suitability of the former. Table \ref{tab:madest} reports the simulated variation and Table \ref{tab:the_se} reports the analytic variation of the 500 estimates of $\hat{\beta}_q$ for $q=0.5$ and $0.75$ using fixed and data-driven tuning constants. Since the results for $q=0.25$ and $q=0.75$ are very similar, only $q=0.75$ is reported. The most efficient method is presented in bold, and also the percent improvement in reduced variation is also presented in comparison to the fixed default tuning constant of 1.345. 

We observe that the data-driven approaches (MQ AV and MQ Inv) outperform the usual fixed tuning constant of 1.345 in most of the settings. The MQ AV improves the efficiency in 80\% of the results for the simulated variation and 70\% for the analytic variation. And the MQ Inv improves the efficiency in 60\% of the results for both measures. The notable settings where efficiency was not improved was for $q=0.75$ with the Normal 5\% and t-distributions where the $c=1.345$ is marginally better for both measures. For the results which did show an improvement the two tables generally show an improvement in efficiency by up to 10\%, for example the Cauchy distribution and $q=0.5$. A comparison of the two tables reveals similar values which suggest that the analytic standard error is a useful estimate, with a large sample size at least.

\begin{table}[ht]
\caption{\label{tab:madest} MAD of $\hat{\bbeta}_q$ for $q=0.5$ and $0.75$ using MQ with tuning constants $c=(1.345, 4)$, MQ AV and MQ Inv. The results are based on 500 Monte-Carlo replications for each scenario. Sample size is $n=10,000$.}
\centering
\resizebox {1\linewidth}{!}{
\begin{tabular}{l|ccccc|ccccc}\hline
                & \multicolumn{3}{c}{Normal distribution}&$t(3)$ & Cauchy& \multicolumn{3}{c}{Normal distribution}&$t(3)$ & Cauchy\\ \hline
Level of contam. & 0\%&5\% &20\% &- & -& 0\%&5\% &20\% &- & -\\ \hline
{{\textbf{$\beta_0$}}} & \multicolumn{5}{c|}{$q=0.5$}&  \multicolumn{5}{c}{$q=0.75$}\\ \hline
   MQ ($c=1.345$) & 0.0132 & 0.0160 & 0.0204 & 0.0194 & 0.0272 & 0.0135 & \bf{0.0174} & 0.0279 & \bf{0.0233} & 0.0387 \\
          MQ AV & \bf{0.0127} & \bf{0.0159} & \bf{0.0198} & \bf{0.0188} & 0.0258 & \bf{0.0127} & 0.0184 & \bf{0.0250} & 0.0250 & \bf{0.0386} \\
          \textit{\% improv.*} & \textit{3.60} &\textit{ 0.82} & \textit{2.97} & \textit{2.83} & \textit{5.18} & \textit{5.88} & \textit{-6.16} & \textit{10.56} & \textit{-7.43} & \textit{0.42} \\
         MQ Inv & 0.0128 & 0.0163 & 0.0202 & 0.0206 & \bf{0.0252} & 0.0128 & 0.0182 & 0.0257 & 0.0234 & 0.0396 \\
         \textit{\% improv.*} & \textit{2.94} & \textit{-1.47} & \textit{1.14} & \textit{-6.49} & \textit{7.35} & \textit{5.55} & \textit{-4.62} & \textit{8.09} & \textit{-0.48} & \textit{-2.16} \\
   MQ ($c=4.0$) & 0.0128 & 0.0205 & 0.0348 & 0.0225 & 0.0397 & 0.0128 & 0.0237 & 0.0598 & 0.0257 & 0.0578 \\
   \hline
\hline
{{\textbf{$\beta_1$}}} & \multicolumn{5}{c|}{$q=0.5$}&  \multicolumn{5}{c}{$q=0.75$}\\ \hline
  MQ ($c=1.345$) & 0.0098 & 0.0112 & 0.0147 & 0.0141 & 0.0203 & 0.0106 & \bf{0.0129} & 0.0192 & 0.0166 & 0.0291 \\
  MQ AV & 0.0093 & 0.0113 & 0.0136 & \bf{0.0138} & 0.0183 & \bf{0.0100} & \bf{0.0129} & 0.0177 & \bf{0.0164} & \bf{0.0282} \\
  \textit{\% improv.*} &  \textit{4.51} & \textit{-0.46} & \textit{7.53} & \textit{2.49} & \textit{9.67} & \textit{6.10} & \textit{-0.29} & \textit{8.05} & \textit{1.34} & \textit{3.10} \\
  MQ Inv & \bf{0.0092} & \bf{0.0111} & \bf{0.0135} & 0.0148 & \bf{0.0177} & \bf{0.0100} & 0.0132 & 0.0186 & 0.0173 & 0.0286 \\
  \textit{\% improv.*} & \textit{5.39} & \textit{0.65} & \textit{8.40} & \textit{-5.01} & \textit{12.70} & \textit{6.16} & \textit{-2.31} & \textit{3.40} & \textit{-4.39} & \textit{1.51} \\ 
  MQ ($c=4.0$) & \bf{0.0092} & 0.0146 & 0.0267 & 0.0151 & 0.0286 & \bf{0.0100} & 0.0168 & 0.0390 & 0.0197 & 0.0404 \\
\hline
\hline
\multicolumn{11}{l|}{* Percentage improvement compared to MQ ($c=1.345$)}
\end{tabular}}
\end{table}

\begin{table}[ht]
\caption{\label{tab:the_se} Median of standard error estimates of $\hat{\bbeta}_q$ for $q=0.5$ and $0.75$ using MQ with tuning constants $c=(1.345, 4)$, MQ AV and MQ Inv. The results are based on 500 Monte-Carlo replications for each scenario. Sample size is $n=10,000$.}
\centering
\resizebox {1\linewidth}{!}{
\begin{tabular}{l|ccccc|ccccc}\hline
                & \multicolumn{3}{c}{Normal distribution}&$t(3)$ & Cauchy& \multicolumn{3}{c}{Normal distribution}&$t(3)$ & Cauchy\\ \hline
Level of contam. & 0\%&5\% &20\% &- & -& 0\%&5\% &20\% &- & -\\
\hline
{{\textbf{$\beta_0$}}} & \multicolumn{5}{c|}{$q=0.5$}&  \multicolumn{5}{c}{$q=0.75$}\\ \hline
  MQ ($c=1.345$) & 0.0145 & \bf{0.0157} & 0.0208 & 0.0178 & 0.0265 & 0.0156 & \bf{0.0176} & 0.0274 & \bf{0.0211} & 0.0373 \\
  MQ AV          & \bf{0.0141} & \bf{0.0157} & \bf{0.0198} & \bf{0.0175} & 0.0230 & \bf{0.0149} & \bf{0.0176} & \bf{0.0251} & 0.0212 & 0.0350 \\
  \textit{\% improv.*} & \textit{2.51} & \textit{-0.21} & \textit{4.91} & \textit{1.78} & \textit{13.18} & \textit{4.28} & \textit{-0.08} & \textit{8.13} & \textit{-0.47} & \textit{6.16} \\
  MQ Inv         & \bf{0.0141} & 0.0160 & 0.0199 & 0.0188 & \bf{0.0225} & \bf{0.0149} & 0.0180 & 0.0257 & 0.0220 & \bf{0.0346} \\
  \textit{\% improv.*} & \textit{2.50} & \textit{-1.60} & \textit{4.37} & \textit{-5.49} & \textit{15.33} & \textit{4.37} & \textit{-2.35} & \textit{6.13} & \textit{-4.45} & \textit{7.03} \\
  MQ ($c=4.0$)   & \bf{0.0141} & 0.0191 & 0.0361 & 0.0208 & 0.0408 & \bf{0.0149} & 0.0231 & 0.0561 & 0.0245 & 0.0587 \\
   \hline
\hline
{{\textbf{$\beta_1$}}} & \multicolumn{5}{c|}{$q=0.5$}&  \multicolumn{5}{c}{$q=0.75$}\\
\hline
  MQ ($c=1.345$) & 0.0103 & \bf{0.0111} & 0.0147 & 0.0126 & 0.0188 & 0.0110 & \bf{0.0124} & 0.0194 & \bf{0.0149} & 0.0264 \\
  MQ AV          & \bf{0.0100} & \bf{0.0111} & \bf{0.0140} & \bf{0.0124} & 0.0163 & \bf{0.0106} & 0.0125 & \bf{0.0178} & 0.0150 & 0.0247 \\
  \textit{\% improv.*} & \textit{2.50} & \textit{-0.21} & \textit{4.91} & \textit{1.84} & \textit{13.20} & \textit{4.22} & \textit{-0.20} & \textit{8.19} & \textit{-0.45} & \textit{6.19} \\
  MQ Inv         & \bf{0.0100} & 0.0113 & 0.0141 & 0.0133 & \bf{0.0159} & \bf{0.0106} & 0.0127 & 0.0182 & 0.0155 & \bf{0.0245} \\
  \textit{\% improv.*} & \textit{2.50} & \textit{-1.56} & \textit{4.35} & \textit{-5.51} & \textit{15.36} & \textit{4.29} & \textit{-2.34} & \textit{6.02} & \textit{-4.40} & \textit{7.10} \\
  MQ ($c=4.0$)   & \bf{0.0100} & 0.0135 & 0.0255 & 0.0147 & 0.0289 & \bf{0.0106} & 0.0163 & 0.0396 & 0.0173 & 0.0416 \\
\hline
\hline
\multicolumn{11}{l|}{* Percentage improvement compared to MQ ($c=1.345$)}
\end{tabular}}
\end{table}

\section{Application: EU-SILC data of Tuscany region (Italy)}\label{sec:cbs}
The application that we consider in this paper is from the 2008 wave of the European Survey on Income and Living Conditions (EU-SILC) conducted by the Italian Office of Statistics (ISTAT) in Italy. The survey allows to produce estimates of living condition indicators at national and regional (NUTS 2) levels. EU-SILC is a rotating panel survey with a 75 per cent overlap of samples in successive years. Each sample is drawn according to a stratified two-stage sample design where municipalities are the primary sampling units (PSUs), while households are the secondary sampling units (SSUs). The PSUs are divided into strata according to their population size; the SSUs are selected by means of systematic sampling in each PSU. 

In this paper, as an example, we consider the EU-SILC data from the Tuscany region. The data consist of 1495 measurements. The outcome of interest is the equivalised income and the reference year is 2007. In this paper the household equivalised income is computed by using the modified OECD scale \citep{Hagenaars:94}: it is calculated for each household as the household total disposable net income divided by the equivalised household size, which gives a weight of 1.0 to the first adult, 0.5 to other persons aged 14 or more and 0.3 to each child aged less than 14. The key explanatory variables are as follows: \texttt{owner}, indicates the ownership of the house and it has two levels (owner or free accommodation/other); \texttt{age}, represents the age of the head of the household; \texttt{work}, is the occupational status of the head of the household and it has two levels (working/other); \texttt{gender}, defines the gender of the head of the household; \texttt{year\_edu}, indicates the years in education of the head of the household; \texttt{hsize} is the size of the household. These variables can be used as covariates in the MQ model for explaining the variability of the equivalised income.

The coefficient estimates for the regression are shown for the MQ with $c=(1.345, 4)$ and the two data-driven approaches, using $q=0.5, 0.75$ and $0.9$ (Table \ref{tab:EUSILC}). Note that optimal $c$ for MQ AV were 1.34, 1.28 and 2.03 respectively for $q=0.5, 0.75$ and $0.9$. For MQ Inv the global optimal was 2.1. The primary finding here is that the differences between the different estimation types is not trivial, with particularly large differences in the MQ ($c=4.0$) estimates. Comparisons between the default MQ ($c=1.345$) and the data-driven approaches are noteworthy, especially as $q$ becomes more extreme. This highlights that there are practical implications to tuning constants and which tuning constant is selected for analysis.

More importantly than the differences in the estimates themselves are the differences in their efficiency. Table~\ref{tab:EUSILC_se} shows the estimated standard errors based on Equation~\eqref{eq:eff} (equivalent to the analytic variation measure in the previous section). The estimation type with the smallest standard error is marked in bold. For all coefficients and all $q$ the smallest standard errors come from the MQ AV. Generally the differences between the MQ AV and MQ ($c=1.345$) estimator are greatest for $q=0.9$ (6.98\% better for MQ AV), and relatively similar for $q=0.5$ and 0.75 (less than 1 \% difference). The standard errors are generally larger for the MQ Inv method ($c=2.1$) compared to the MQ ($c=1.345$) estimator except for when $q=0.9$. However, it is still much better compared to the MQ ($c=4$) estimates. This suggests that the global approach might favour efficiency in the tails for extreme $q$, over $q$ closer to 0.5.

\begin{table}[ht]
\caption{\label{tab:EUSILC} EU-SILC estimates of $\bbeta_q$ for $q=0.5, 0.75$ and $0.9$ using MQ with tuning constants $c=(1.345, 4)$, MQ AV and MQ Inv.}
\centering
\begin{tabular}{l|ccccccc}
  \hline
 & Intercept & owner & age & work & gender & year\_edu & hsize \\
  \hline
   & & & &$q=0.5$ & & & \\
  \hline
  MQ ($c=1.345$) & 2434.58 & 3175.02 & 25.66 & 4604.95 & 1681.68 & 547.45 & 531.79 \\
  MQ AV & 2442.47 & 3173.89 & 25.59 & 4600.31 & 1681.35 & 547.33 & 531.75 \\
  MQ Inv & 1427.37 & 3391.64 & 35.75 & 5000.12 & 1786.89 & 582.31 & 495.95 \\
  MQ ($c=4$) & 40.09 & 3216.53 & 53.07 & 5239.02 & 2231.33 & 702.93 & 335.19 \\
    \hline
     & & & & $q=0.75$& & & \\
  \hline
  MQ ($c=1.345$) & 2116.57 & 4037.44 & 50.69 & 4494.24 & 2489.72 & 800.96 & 354.56 \\
  MQ AV & 2156.79 & 4064.07 & 49.83 & 4470.87 & 2478.46 & 798.35 & 366.33 \\
  MQ Inv & 1503.12 & 3863.19 & 61.87 & 4967.26 & 2574.13 & 824.88 & 245.71 \\
  MQ ($c=4$) & -229.47 & 3262.23 & 88.94 & 5487.77 & 3186.48 & 970.54 & 20.53 \\
    \hline
     & & & & $q=0.9$& & & \\
  \hline
  MQ ($c=1.345$) & -511.47 & 4373.92 & 120.35 & 4977.89 & 3622.58 & 1256.91 & -97.48 \\
  MQ AV & -167.33 & 3533.45 & 122.51 & 5116.03 & 3816.77 & 1263.15 & -169.00 \\
  MQ Inv & -170.65 & 3463.26 & 123.39 & 5142.62 & 3845.07 & 1262.02 & -173.77 \\
  MQ ($c=4$) & -1260.86 & 2247.87 & 155.97 & 5919.82 & 4902.82 & 1378.98 & -541.47 \\
   \hline
\end{tabular}
\end{table}

\begin{table}[ht]
\caption{\label{tab:EUSILC_se} EU-SILC standard error estimates of $\hat{\bbeta}_q$ for $q=0.5, 0.75$ and $0.9$ using MQ with tuning constants $c=(1.345, 4)$, MQ AV and MQ Inv.}
\centering
\begin{tabular}{l|ccccccc}
  \hline
 & Intercept & owner & age & work & gender & year\_edu & hsize \\
  \hline
   & & & &$q=0.5$ & & & \\
  \hline
  MQ ($c=1.345$) & 1364.84 & 590.20 & 13.93 & 666.11 & 485.75 & 49.37 & 182.60 \\
  MQ AV          & \bf{1362.41} & \textbf{589.15} & \textbf{13.90} & \textbf{664.93} & \textbf{484.89} & \textbf{49.28} & \textbf{182.28} \\
  MQ Inv         & 1438.18 & 621.91 & 14.68 & 701.91 & 511.85 & 52.03 & 192.41 \\
  MQ ($c=4$)     & 1707.93 & 738.56 & 17.43 & 833.56 & 607.86 & 61.78 & 228.50 \\
  \hline
     & & & & $q=0.75$& & & \\
  \hline
  MQ ($c=1.345$) & 1874.42 & 810.56 & 19.13 & 914.82 & 667.11 & 67.81 & 250.78 \\
  MQ AV          & \textbf{1873.11} & \textbf{809.99} & \textbf{19.12} & \textbf{914.18} & \textbf{666.65} & \textbf{67.76} & \textbf{250.60} \\
  MQ Inv         & 1976.51 & 854.71 & 20.17 & 964.64 & 703.45 & 71.50 & 264.44 \\
  MQ ($c=4$)     & 2557.01 & 1105.73 & 26.10 & 1247.96 & 910.05 & 92.50 & 342.10 \\
  \hline
     & & & & $q=0.9$ & & & \\
  \hline
  MQ ($c=1.345$) & 3647.05 & 1577.10 & 37.22 & 1779.95 & 1298.00 & 131.93 & 487.93 \\
  MQ AV         & \textbf{3392.37} & \textbf{1466.97} & \textbf{34.62} & \textbf{1655.66} & \textbf{1207.36} & \textbf{122.72} & \textbf{453.86} \\
  MQ Inv        & 3421.11 & 1479.40 & 34.91 & 1669.68 & 1217.59 & 123.76 & 457.71 \\
  MQ ($c=4$)   & 4587.28 & 1983.69 & 46.82 & 2238.84 & 1632.63 & 165.94 & 613.73 \\
   \hline
\end{tabular}
\end{table}

\section{Summary}\label{sec:final}
In the first part of this paper we showed that two previously proposed MQ scale estimators, nMAD and ML estimators, are generally not appropriate for MQ estimation. Hence we suggest that the widely accepted nMAD approach cease to be used in MQ models. Further, the maximum likelihood approach to MQ regression has also been shown to be non-robust as well as unsuitable with extreme $q$. The proposed MM scale estimator was shown to perform relatively well, and similarly to the cMAD estimator. We propose that either of these two estimators would be suitable in practice, however advise the use of the cMAD approach merely because of the relative simplicity in concept and implementation.

With an understanding of the best scale estimator, two data-driven methods for selecting an optimal tuning constant were introduced. The first based on a direct extension of \citet{Wan07} (MQ AV) and the second using a novel approach based on an inverse MQ function (MQ Inv). Both methods offered improvements to the efficiency of the MQ estimator compared to the typical choice of $c=1.345$, albeit with some exceptions in some settings. Each of the two methods address the problem of subjective tuning constant choices by providing optimal alternatives based on the respective criteria. The MQ AV method offers optimal tuning constants for a given $q$ based on minimising asymptotic variance. With smaller sample sizes this method may provide lower than necessary tuning constants when no outliers are present, but generally performs well over heavy-tailed distributions. The global approach of the MQ Inv provides one single tuning constant to be used when fitting an ensemble of MQ models, which is often the case. This global optimal tuning constant minimises the effect of contamination to an otherwise normal distribution. It was shown that this method did not perform as well when the residuals were not similar to a contaminated normal distribution. Further work could adapt the MQ Inv method beyond just the contaminated normal assumption.

Further to alleviating the problem of subjective tuning constant selection, these two methods also provide a diagnostic tool to assess the level of robustness required for efficient estimation. This could be especially useful when determining whether robust methods should be used over other approaches, for example if deciding whether expectile regression is appropriate across all $q$ for a certain data set with possible outliers.

These findings regarding the MQ scale estimator and tuning constant provide a necessary insight into two overlooked areas of MQ regression. They give insight into improved approaches to future MQ applications as well as a useful and novel framework for selecting data-driven tuning constants rather than merely accepting a default choice to be appropriate.

\pagebreak
\bibliographystyle{chicago}
\bibliography{Datenbank}


\appendix

\section{Appendix}\label{sec:appendix}
From \citet{Jon94}, the inverse MQ function can be found by rearranging the functional given in Equation~\eqref{eq:MQfunct} to:
\begin{align}
G_c(x;F) &= \frac{ \int_{-\infty}^{x} \psi_c\left(\frac{y-x}{\sigma_{q}}\right) dF(y)}{ \int_{-\infty}^{x} \psi_c\left(\frac{y-x}{\sigma_{q}}\right) dF(y) - \int_{x}^{\infty} \psi_c\left(\frac{y-x}{\sigma_{q}}\right) dF(y)}.\label{eq:Gq1}
 \end{align}
This function is comprised of two unique definite integrals which can be expanded like so:
\begin{align*}
         &\int_{-\infty}^{x} \psi_c \left(\frac{y-x}{\sigma_{q}}\right) dF(y) \\
         =& \int_{-\infty}^{x-c\sigma_{q}} -c dF(y) + \int_{x-c\sigma_{q}}^{x}  \left(\frac{y-x}{\sigma_{q}}\right) dF(y)\\
           =& -c \int_{-\infty}^{x-c\sigma_{q}} dF(y) + \frac{1}{\sigma_{q}} \left[\int_{x-c\sigma_{q}}^{x}  y dF(y) - x\int_{x-c\sigma_{q}}^{x}   dF(y) \right] \\
            =&  -c F(x-c\sigma_{q}) + \frac{1}{\sigma_{q}} \left( \left[H(x)-H(x-c\sigma_{q})\right] 
            - x \left[F(x)-F(x-c\sigma_{q})\right] \right) \\
             =& \left(\frac{x}{\sigma_{q}}-c\right) F(x-c\sigma_{q}) + \frac{1}{\sigma_{q}}\left[H(x)-H(x-c\sigma_{q})- x F(x)\right]
\end{align*}
 and,
\begin{align*}
        &\int_{x}^{\infty} \psi_c \left(\frac{y-x}{\sigma_{q}}\right) dF(y) \\
        =&\int_{x}^{x+c\sigma_{q}}  \left(\frac{y-x}{\sigma_{q}}\right) dF(y) + \int_{x+c\sigma_{q}}^{\infty} c dF(y) \\
          =& \frac{1}{\sigma_{q}} \left[ \int_{x}^{x+c\sigma_{q}}  y dF(y) - x \int_{x}^{x+c\sigma_{q}}  dF(y) \right] + c \int_{x+c\sigma_{q}}^{\infty} dF(y) \\
            =& \frac{1}{\sigma_{q}} \left[\left(H(x+c\sigma_{q})-H(x) \right) - x \left(F(x+c\sigma_{q})- F(x) \right) \right] + c \left[ 1 - F(x+c\sigma_{q}) \right] \\
             =& \frac{1}{\sigma_{q}} \left[H(x+c\sigma_{q}) - H(x)  + x F(x) \right]- \left(\frac{x}{\sigma_{q}} + c\right)F(x+c\sigma_{q}) + c
\end{align*}
where $H(t) = \int_{-\infty}^{t} y dF(y)$ is the partial expectation. These two integrals can then be substituted into Equation~\ref{eq:Gq1} to get the inverse Huber $M$-quantile function:
\begin{equation*}
 G_c(x;F) = \frac{ b^- F_y(a^-) + \frac{1}{\sigma_{q}}\left[H(x)-H(a^-)- x F_y(x)\right]}  {\frac{1}{\sigma_{q}}\left[2H(x)-H(a^-)-H(a^+)- 2x F_y(x)\right] + b^- F_y(a^-) +  b^+ F_y(a^+) - c}
\end{equation*}
where $a^- = x-c\sigma_{q}$, $a^+ = x+c\sigma_{q}$, $b^-=\frac{x}{\sigma_{q}}-c$, and $b^+=\frac{x}{\sigma_{q}}+c$.

\end{document}